\documentclass{article}
\usepackage[T1]{fontenc}
\usepackage{times}
\usepackage{subfigure}
\usepackage[colorlinks=true, linkcolor=blue, citecolor=cyan, urlcolor=blue]{hyperref}
\usepackage{cite}
\usepackage{amsmath}
\usepackage{geometry}
\usepackage{graphicx}
\usepackage{float}
\usepackage{indentfirst}
\title{Observational and Theoretical Constraints on \\First-Order Phase Transitions in Neutron Stars}

\author{Zuhua Ji, Jiarui Chen, Gaojian Wu\\
	Department of physics, School of Physical and Mathematical Sciences, \\Nanjing Tech University, Nanjing 210009, PR China}
\date{}
\begin{document}
	\maketitle
	
\begin{abstract}

	Understanding the equation of state (EOS) of neutron stars (NSs) is a fundamental challenge in astrophysics and nuclear physics. A first-order phase transition (FOPT) at high densities could lead to the formation of a quark core, significantly affecting NS properties. This review explores observational and theoretical constraints on such transitions from observational effects. X-ray observations, including mass-radius measurements from NICER and spectral features like quasi-periodic oscillations (QPOs) and cyclotron resonance scattering features (CRSFs), provide indirect evidence of EOS modifications. Gravitational wave detections, particularly from binary NS mergers GW170817, constrain tidal deformability and post-merger oscillations, which may carry signatures of phase transitions. Pulsar timing offers additional constraints through measurements of mass, spin evolution, and glitches, with millisecond pulsars exceeding $2M_{\odot}$ posing challenges to purely hadronic EOSs. Theoretical models and numerical simulations predict that an FOPT could impact gravitational wave signals, twin-star configurations, and NS cooling. Future advancements, including next-generation gravitational wave detectors, high-precision X-ray telescopes, and improved theoretical modeling, will enhance our ability to probe phase transitions in NSs. A combination of these approaches will provide crucial insights into the existence and properties of deconfined quark matter in NS interiors.
	
\end{abstract}

\section{Introduction}

Neutron stars (NSs) are among the most extreme astrophysical objects, with central densities exceeding several times the nuclear saturation density (\(\rho_0\))\cite{ref1}. Understanding their equation of state (EOS) has been a fundamental challenge in astrophysics and nuclear physics in recent years. One key question is whether a first-order phase transition (FOPT) occurs inside NSs, leading to the formation of deconfined quark matter at high densities. Such a transition would manifest as a discontinuity in the pressure-density relation, introducing a latent heat and altering key observables such as the maximum mass, radius, and tidal deformability\cite{ref2}.

Quantum chromodynamics (QCD) predicts that hadronic matter may transition to quark matter at extreme densities\cite{ref3}, with the nature of this transition remaining uncertain. If the transition is first-order, the resulting softening of the EOS could significantly affect NS structure and evolution. Several theoretical models, including the MIT bag model and Nambu–Jona-Lasinio models, suggest that such a transition might occur at densities \(\sim 2-3 \rho_0\)\cite{ref4}, leading to a hybrid star with a hadronic envelope and a quark matter core. The presence of such an FOPT would have significant astrophysical consequences, influencing NS cooling rates, pulsar spin evolution, and gravitational wave (GW) signatures\cite{ref5}. Various constraints are provided by observational effects. X-ray measurements from NICER and other telescopes provide mass-radius constraints, which can help identify deviations from purely hadronic EOSs\cite{ref6}. In X-ray binaries\cite{ref5}, quasi-periodic oscillations (QPOs), cyclotron resonance scattering features (CRSFs), and cooling curves offer indirect insights into the internal composition of NSs. Meanwhile, gravitational waves from binary NS mergers encode information about the EOS, particularly through tidal deformability (\(\Lambda\)) in the inspiral phase and post-merger oscillation frequencies (\( f_{\text{peak}} \)). The GW170817 event has already provided constraints on NS radii, while the GW190814 event raises questions about the nature of a 2.6\(M_{\odot}\) compact object, potentially requiring an exotic EOS\cite{ref7}.

Pulsar timing also offers valuable constraints. The discovery of NSs exceeding \(2M_{\odot}\), such as PSR J0740+6620, suggests that the EOS must remain sufficiently stiff to support high-mass stars, while spin evolution and glitch behavior may reflect changes in the NS core\cite{ref8}. Pulsar timing arrays (PTAs)\cite{ref7} like NANOGrav and SKA are also capable of detecting stochastic GW backgrounds, which could contain imprints of neutron star mergers influenced by phase transitions.

Despite recent advances, the presence and properties of an FOPT in NSs remain uncertain. This review synthesizes current theoretical models, observational evidence, and numerical simulations, discussing how X-ray, GW, and pulsar observations can jointly constrain the existence of a phase transition. It also highlights future observational prospects, including next-generation GW detectors such as the Einstein Telescope (ET) and Cosmic Explorer (CE), which will improve sensitivity to post-merger signals, and advanced X-ray missions like Athena, which will refine mass-radius measurements. Through a multi-messenger approach, these upcoming observations may provide definitive evidence for or against the presence of deconfined quark matter in neutron stars.

The main goal of this review is to explore the FOPT in neutron stars and its observational signatures. In Sec. \ref{2}, we introduce various theoretical models for dense matter, including purely hadronic, hybrid, and quark matter equations of state, and examine how different construction methods such as the Maxwell and Gibbs formalisms impact neutron star structure. In Sec. \ref{3}, we focus on gravitational wave observations from binary neutron star mergers, highlighting how tidal deformability and post-merger oscillations can be used to constrain the EOS and reveal possible phase transitions. Sec. \ref{4} is devoted to pulsar timing observations, where mass measurements, spin evolution, and glitch events are discussed as probes of internal composition and structural changes. in Sec. \ref{5}, we present future prospects, including the role of next-generation gravitational wave and X-ray observatories, as well as upcoming improvements in theoretical modeling and numerical simulations, all of which are expected to significantly advance our understanding of deconfined quark matter in neutron star interiors. Conclusions are drawn in Sec. \ref{6}.

\section{Theoretical Models of First-Order Phase Transitions in Neutron Stars}
\label{2}
The existence of a first-order phase transition (FOPT) in neutron stars, leading to the formation of a deconfined quark core or a mixed-phase region, is a central question in nuclear astrophysics. Theoretical models of dense matter aim to predict the equation of state (EOS) that governs the macroscopic properties of neutron stars, including their mass-radius relation, maximum mass, and tidal deformability. Recent advances in numerical simulations have also provided critical insights into how a phase transition manifests in gravitational wave signals, pulsar timing, and X-ray observations.

Theoretical models of neutron star EOSs generally fall into three categories: purely hadronic models, hybrid models with a hadron-quark phase transition, and models with full deconfinement into quark matter at high densities. Purely hadronic models, such as APR and SLy\cite{ref9}, predict that neutron stars remain composed of nucleons, hyperons, and leptons up to the highest densities before collapsing into a black hole. In contrast, hybrid EOSs introduce an FOPT, typically modeled using the Maxwell construction, where a sharp transition occurs at a critical density, or the Gibbs construction, where a mixed phase forms over a range of densities. Some models, such as the MIT bag model or NJL-based quark matter EOSs, predict a complete transition to quark matter in the core, forming a strange star rather than a conventional neutron star\cite{ref10}.

While the precise nature of the phase transition and its observable consequences remain uncertain, recent numerical simulations offer compelling insights into how such transitions might manifest during neutron star mergers. One of the most direct indicators is the evolution of the central density during and after the coalescence process. If the equation of state includes a first-order phase transition, the remnant may undergo a sharp increase in central density that crosses the critical threshold for deconfinement\cite{ref11}. 

\begin{figure}[H]
	\centering
	\includegraphics[width=0.6\textwidth]{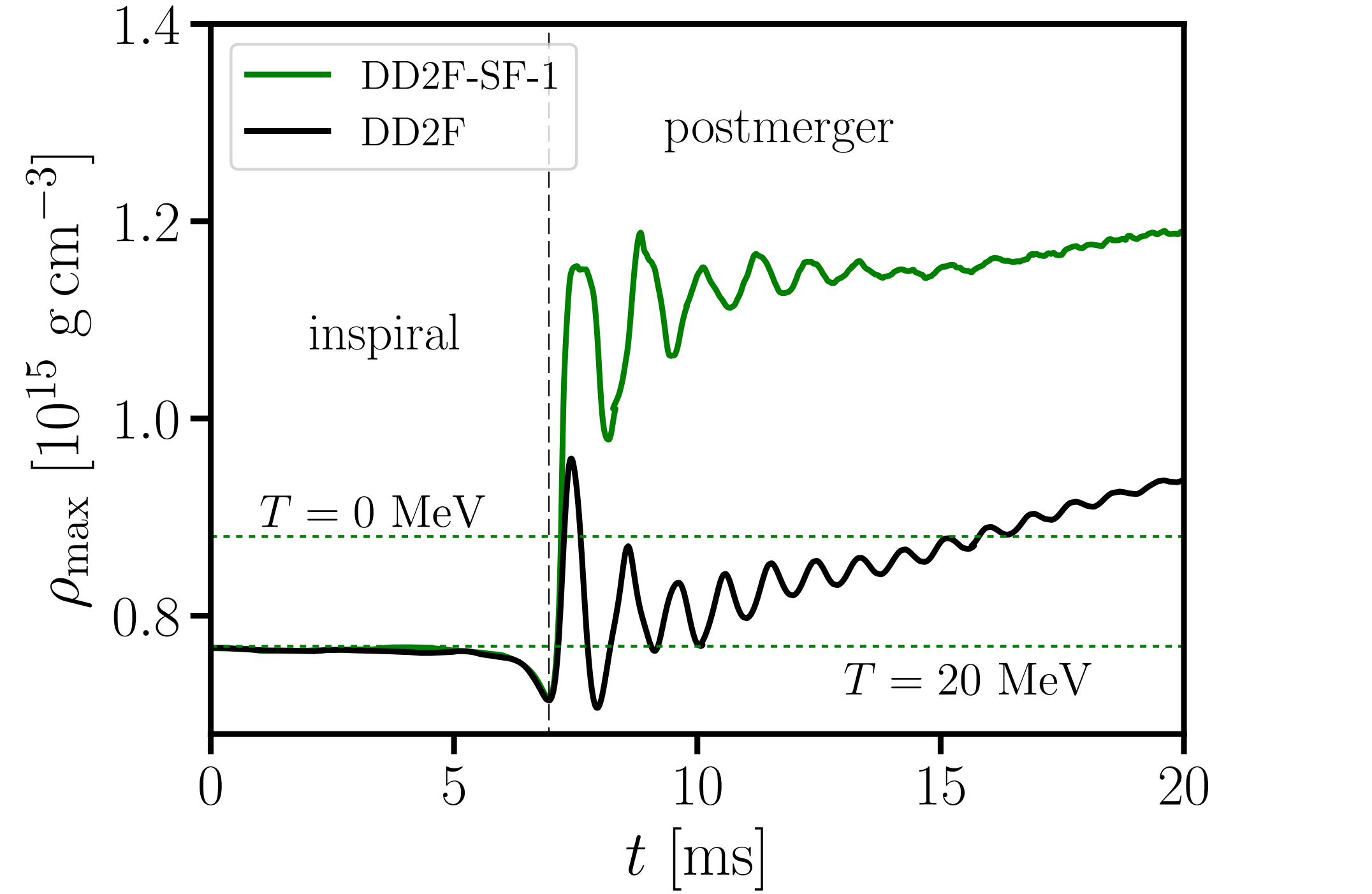}
	\caption{Comparison of the time evolution of maximum rest-mass density in neutron star mergers with and without FOPT.
	}
	\label{fig1}
\end{figure}

This process can significantly alter the post-merger dynamics and gravitational wave signatures, as illustrated in Figure~\ref{fig1} taken from \cite{ref1}. Evolution of the maximum rest-mass density $\rho_{\mathrm{max}}$ during a binary neutron star merger, comparing two equations of state (EOS): the purely hadronic DD2F model (black curve) and the hybrid DD2F-SF-1 model with a first-order phase transition (green curve). 
The horizontal dashed lines indicate the onset densities of the phase transition at $T=0$ and $T=20$ MeV. 
The DD2F-SF-1 model shows a rapid increase in central density post-merger, crossing the phase transition threshold and leading to significant deviations in remnant dynamics. 
This behavior may result in suppressed post-merger gravitational wave emission and early black hole formation, providing a potential signature of deconfined quark matter.

The Maxwell and Gibbs constructions result in distinct observational signatures. A Maxwell-type phase transition leads to a sharp density jump, which can create a twin-star configuration, where stars of similar masses have drastically different radii depending on whether they contain a quark core. This effect has been explored in relativistic mean-field (RMF) models, which show that the transition density and latent heat determine whether a twin-star sequence is observationally viable.In relativistic mean-field (RMF) models. Nucleon interactions in these models are mediated by meson exchange, and the energy density and pressure are determined by the self-consistent solution of the mean-field equations. The equation governing pressure in RMF models is given by

\begin{equation}
	P = \sum_i P_i + \frac{1}{2} m_{\sigma}^2 \sigma^2 - \frac{1}{2} m_{\omega}^2 \omega^2 - \frac{1}{2} m_{\rho}^2 \rho^2
\end{equation}
 The Gibbs construction, by contrast, produces a more gradual softening of the EOS, potentially leading to a suppression of gravitational wave post-merger signals due to rapid collapse\cite{ref12}.

Numerical simulations of neutron star mergers incorporating phase transitions have provided valuable insights into the dynamics of high-density matter. General relativistic hydrodynamics simulations indicate that an FOPT can alter the inspiral waveform by affecting the tidal deformability parameter \(\Lambda\), leading to deviations from purely hadronic EOS predictions. If a phase transition occurs in the remnant neutron star, it can induce a delayed collapse to a black hole, significantly shortening the post-merger lifetime. Studies using the ALF2 and DD2F-SF EOSs show that a strong FOPT can reduce the dominant post-merger gravitational wave frequency \( f_{\text{peak}} \) by up to 300 Hz, providing a detectable signature in next-generation detectors.

In quark-hadron crossover (QHC) models, the transition between hadronic and quark matter is gradual rather than abrupt. These models are motivated by lattice QCD simulations, which suggest that chiral symmetry restoration in QCD can occur as a smooth crossover rather than a sharp phase transition. The crossover is typically parameterized using a weighted pressure function\cite{ref8}:

\begin{equation}
	P_{\text{QHC}}(\epsilon) = w P_{\text{hadronic}} + (1-w) P_{\text{quark}}
\end{equation}
where \(w(\rho)\) smoothly interpolates between the two phases.

Theoretical uncertainties remain in modeling the hadron-quark transition, particularly in the treatment of the speed of sound \( c_s \) in dense matter. In purely hadronic EOSs, \( c_s^2 \) typically remains below 0.5, while some quark matter models predict values approaching the conformal limit of \( c_s^2 = 1/3 \). Hybrid EOSs with FOPTs often require an intermediate stiffening to support neutron stars above \(2M_{\odot}\), with models such as the quarkyonic matter EOS suggesting that a peak in \( c_s^2 \) may occur before full deconfinement.

The viability of different first-order phase transition (FOPT) models can be further assessed by examining their predictions in the mass-radius ($M$–$R$) diagram under theoretical and observational constraints. 
In particular, the speed of sound must not exceed the speed of light ($c_s < c$), which imposes a causality limit on the equation of state (EOS). 
At the same time, mass and radius measurements from NICER and gravitational wave events such as GW170817 provide empirical constraints that any realistic EOS must satisfy. 
To explore how varying the phase transition onset energy density $\epsilon_{\mathrm{tr}}$ influences EOS predictions, we compare their resulting $M$–$R$ curves with both causality and astrophysical constraints, as shown in Figure~\ref{2} taken from \cite{ref11}.

\begin{figure}[H]
	\centering
	\includegraphics[width=0.6\textwidth]{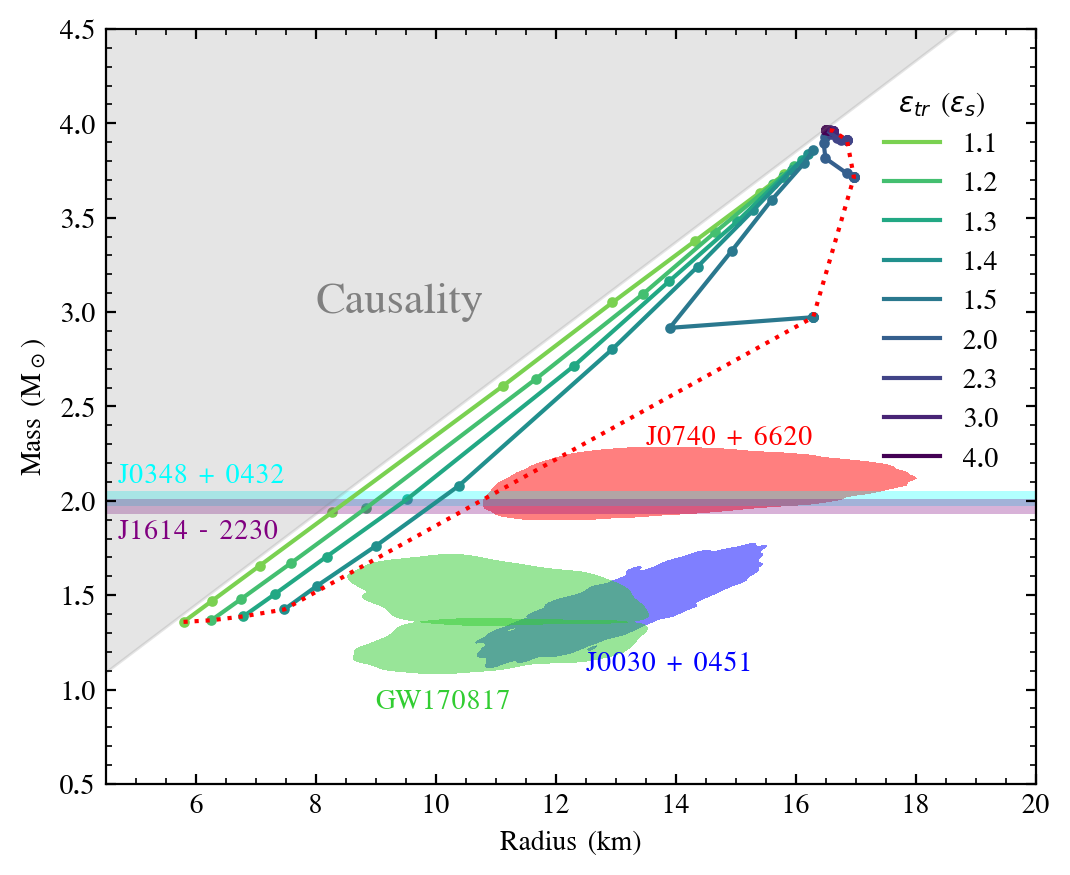}
	\caption{Mass-radius relations under causality constraints for different values of the phase transition onset energy density $\epsilon_{\mathrm{tr}}$, derived by solving the three-parameter scale-free Tolman–Oppenheimer–Volkoff (TOV) equations with $\epsilon_1$ fixed at 150 MeV. 
		Each colored line represents a set of EOSs with different $\epsilon_{\mathrm{tr}}$, and the gray shaded area marks the region where causality (i.e., $c_s > c$) is violated. 
		Scatter points indicate the maximum mass $M_{\mathrm{TOV}}$ for each EOS. 
		Colored contours show astrophysical constraints from observations: 
		red—PSR J0740$+$6620, blue—PSR J0030$+$0451, green—GW170817, cyan—PSR J0348$+$0432, and purple—PSR J1614$-$2230. 
		These constraints help evaluate the viability of different EOSs with respect to both causality and observational limits.
	}
	\label{fig2}
\end{figure}

One of the key challenges in simulating phase transitions in neutron stars is the need for high-resolution, multi-scale modeling. Finite-temperature effects, which become relevant in neutron star mergers, introduce additional complexities in determining the EOS at high densities. Recent advances in lattice QCD at nonzero baryon density and Bayesian statistical inference using gravitational wave and X-ray data have provided improved constraints on the possible phase transition region. Future work integrating multi-messenger observations with theoretical modeling will be crucial in determining whether neutron stars undergo a first-order phase transition and, if so, at what density this transition occurs.

\section{Gravitational Wave Observations}
\label{3}
Observations from both X-ray and gravitational wave (GW) astronomy provide complementary constraints on the neutron star equation of state (EOS), particularly under conditions where a first-order phase transition (FOPT) to quark matter may occur. While X-ray data shed light on surface and inner structure properties via thermal emissions, cyclotron lines, and QPOs, gravitational wave signals probe bulk and dynamical properties such as tidal deformability, post-merger oscillations, and maximum mass constraints. This section presents a unified discussion, grouped by the physical quantities constrained.

\subsection{Radius and Mass}

Quasi-periodic oscillations (QPOs) in accreting neutron stars arise from interactions between the inner accretion disk and the neutron star’s gravitational field, offering indirect constraints on the mass–radius relation. Systems such as 4U 1636-53 and 4U 1820-30 provide valuable observational evidence in this regard. 4U 1636-53, located in the Galactic disk, contains a neutron star accreting matter from a low-mass main-sequence companion, while 4U 1820-30, situated in the globular cluster NGC 6624, features a white dwarf companion and an ultra-short orbital period of about 11 minutes. Both systems exhibit prominent kilohertz QPOs (kHz QPOs), which are linked to the motion of matter near the innermost stable circular orbit (ISCO), where relativistic effects dominate. If a first-order phase transition causes a sudden contraction of the neutron star radius, the ISCO would shift inward, thereby altering the QPO frequency distribution.

The relativistic precession model (RPM) describes how QPO frequencies are connected to the Keplerian and epicyclic motions of disk material. The lower (\(\nu_L\)) and upper (\(\nu_U\)) QPO frequencies satisfy  

\begin{equation}
	\nu_L = \nu_K - \nu_r, \quad \nu_U = \nu_K
\end{equation}
where \(\nu_K\) is the Keplerian frequency, and \(\nu_r\) is the radial epicyclic frequency\cite{ref13}. A phase transition could modify \(\nu_K\) by changing the neutron star's compactness, leading to detectable deviations in these frequency relations. Observations of QPO frequency shifts in sources like Sco X-1 and GX 339-4 have been interpreted as potential evidence for a softening of the EOS at high densities, consistent with the formation of deconfined quark matter.

Gravitational waves provide complementary constraints via the tidal deformability parameter \(\Lambda\)\cite{ref11}, which is sensitive to radius and EOS stiffness:

\begin{equation}
	\Lambda = \frac{2}{3} k_2 \left(\frac{c^2 R}{G M}\right)^5
\end{equation}

The LIGO-Virgo detection of GW170817 constrained the combined tidal deformability \(\tilde{\Lambda}\) of the merging neutron stars to be  

\begin{equation}
	70 \leq \tilde{\Lambda} \leq 720
\end{equation}

with the most probable values between 300 and 400, favoring a relatively soft EOS. If a FOPT occurs at supranuclear densities, it could significantly reduce \(\Lambda\), making neutron stars more compact. The comparative distinctions across different models\cite{ref16} are presented in Table \ref{tab:tidal_deformability}.

\begin{table}[h]
	\centering
	\caption{Tidal Deformability Predictions for Different EOS Models}
	\begin{tabular}{lccc}
		\hline
		EOS Model & Phase Transition & $R_{1.4}$ (km) & $\Lambda_{1.4}$ \\
		\hline
		APR4      & No             & 11.0  & 250  \\
		SLy       & No             & 11.8  & 280  \\
		H4        & No             & 13.6  & 800  \\
		ALF2      & Yes (Quark)    & 12.4  & 420  \\
		DD2F-SF   & Yes (Hybrid)   & 11.5  & 150  \\
		MIT-Bag   & Yes (Quark)    & 10.2  & 50   \\
		\hline
	\end{tabular}
	\label{tab:tidal_deformability}
\end{table}

EOS models with phase transitions tend to predict lower \(\Lambda_{1.4}\) and smaller radii, supporting the hypothesis that a softening of the EOS (e.g., due to quark matter) leads to more compact stars. These predictions align well with LIGO constraints.

The detection of GW190814, which involved a 2.6\(M_{\odot}\) compact object merging with a 23\(M_{\odot}\) black hole, has sparked debates about whether such a massive object could be a neutron star. If it were, its existence would challenge current hadronic EOS models, as most predict a maximum mass around 2.1–2.3\(M_{\odot}\). However, EOS models incorporating a FOPT, such as those with quark cores or hybrid stars, could support ultra-massive neutron stars up to 2.6\(M_{\odot}\).

The speed of sound \( c_s \) in neutron star matter plays a critical role in determining the maximum mass. If an FOPT occurs, the EOS must include a region where \( c_s^2 \) is non-monotonic, featuring a stiffening effect at extreme densities:

\begin{equation}
	c_s^2 = \frac{\partial P}{\partial \epsilon}
\end{equation}

Models like DD2F-SF predict that a brief increase in \( c_s^2 \) following an FOPT could allow a neutron star to remain stable at 2.6\(M_{\odot}\), potentially explaining GW190814's secondary object.

\begin{figure}[H]
	\centering
	\includegraphics[width=0.8\textwidth]{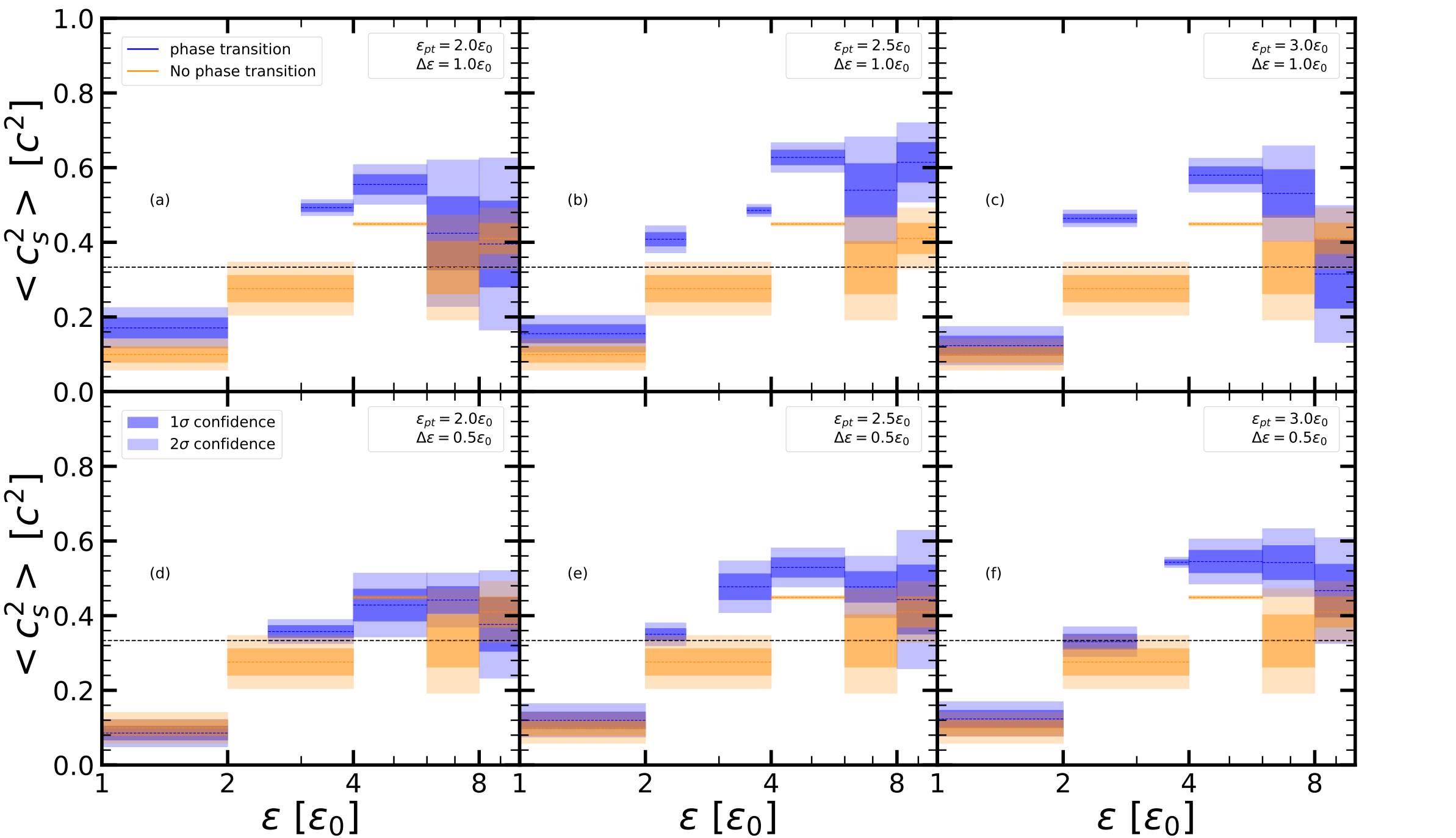}
	\caption{Average squared sound speeds $\langle c_s^2 \rangle$ predicted by deep neural networks (DNNs) for scenarios with (blue) and without (orange) a first-order phase transition, under various combinations of transition energy density $\varepsilon_{\mathrm{pt}}$ and latent heat $\Delta\varepsilon$. 
		The six panels correspond to different parameter sets, as indicated. 
		Shaded bands represent the $1\sigma$ (dark color) and $2\sigma$ (light color) confidence intervals. 
		The horizontal dashed line denotes the conformal limit $\langle c_s^2 \rangle = 1/3$. 
		These results reflect the influence of phase transition parameters on the stiffness of the equation of state.(Figure is taken from \cite{ref9})
	}
	\label{fig3}
\end{figure}

As illustrated in Fig.~\ref{fig3}, the average squared speed of sound $\langle c_s^2 \rangle$ in neutron star matter is evaluated across different energy density intervals, using deep neural networks (DNNs) trained on models with and without a first-order phase transition. In the absence of a phase transition, $\langle c_s^2 \rangle$ increases gradually and peaks around $6\varepsilon_0$, before decreasing toward the high-density regime ($6\varepsilon_0$ to $8\varepsilon_0$), approaching the conformal bound $c_s^2 = 1/3$. Across the entire density domain, the sound speed remains below 0.5. In contrast, the inclusion of a phase transition results in a systematically higher sound speed, indicating a stiffer equation of state. Notably, the maximum value of $\langle c_s^2 \rangle$—including contributions from the quark phase—can reach approximately 0.7.

\subsection{Compactness and Gravitational Redshift}

Cyclotron resonance scattering features (CRSFs)\cite{ref12} provide another means of probing neutron star compactness by measuring the interaction of X-ray photons with electrons in the strong magnetic fields near the stellar surface. These features, detected in pulsars such as Her X-1, Vela X-1, and 4U 1538-52, appear as absorption lines at characteristic energies  

\begin{equation}
	E_c = \frac{\hbar e B}{m_e c}
\end{equation}

The observed energy is redshifted due to the gravitational field of the neutron star, given by  

\begin{equation}
	E_{\text{obs}} = \frac{E_c}{1 + z}, \quad 1 + z = \left( 1 - \frac{2GM}{Rc^2} \right)^{-1/2}
\end{equation}

A phase transition leading to a quark core would reduce \(R\) while keeping \(M\) constant, increasing \(z\) and shifting the CRSF energy downward. Alternatively, if the transition results in a mixed-phase region that expands the star’s radius, the redshift decreases, shifting CRSF lines upward.

Observations by NuSTAR and Insight-HXMT have revealed systematic shifts in CRSF energies that deviate from standard dipole field models. These variations may be linked to changes in the EOS, with certain anomalies suggesting a transition from a stiff hadronic phase to a softer quark matter phase. In particular, transient CRSF shifts in accreting pulsars like V 0332+53 have been observed across different luminosity states, which could indicate structural changes in the neutron star due to a density-driven phase transition. Future high-resolution X-ray spectroscopy will be crucial in determining whether these spectral variations are linked to EOS modifications and the existence of a first-order phase transition in neutron stars.

\subsection{Post-Merger Dynamics}

Following the inspiral phase, the post-merger remnant exhibits characteristic oscillation frequencies (\( f_{\text{peak}} \)), which provide valuable information about the high-density EOS\cite{ref15}. These frequencies depend on the stiffness of the EOS and whether a phase transition occurs during the merger process. In a purely hadronic EOS, the remnant neutron star remains stable for a longer duration, producing a prolonged GW signal. However, if a FOPT occurs at high densities, the remnant may undergo rapid collapse into a black hole due to loss of pressure support.

\begin{equation}
	f_{\text{peak}} \approx 2.5 - 4.0 \text{ kHz}
\end{equation}

\begin{table}[h]
	\centering
	\caption{Post-Merger Oscillation Frequencies and Collapse Times for Various EOS Models}
	\begin{tabular}{lccc}
		\hline
		EOS Model  & $ f_{\text{peak}} $ (kHz) & Collapse Time (ms) & Phase Transition? \\
		\hline
		APR4       & 3.2             & > 20 ms         & No             \\
		SLy        & 3.5             & > 10 ms         & No             \\
		H4         & 2.8             & > 30 ms         & No             \\
		ALF2       & 2.9             & ~5 ms           & Yes (Quark)    \\
		DD2F-SF    & 3.0             & < 3 ms          & Yes (Hybrid)   \\
		MIT-Bag    & 2.5             & < 1 ms          & Yes (Quark)    \\
		\hline
	\end{tabular}
	\label{tab:post_merger}
\end{table}

EOS models with phase transitions (e.g., MIT-Bag, DD2F-SF) lead to significantly faster collapse times and lower oscillation frequencies, potentially suppressing the post-merger GW signal. The absence of a strong post-merger signal in GW170817 is consistent with this behavior.

In order to better understand how a first-order phase transition (FOPT) modifies the post-merger gravitational wave spectrum, it is instructive to examine the relationship between the dominant post-merger frequency $f_{\mathrm{peak}}$ and the tidal deformability $\Lambda_{1.35}$. 

\begin{figure}[H]
	\centering
	\includegraphics[width=0.6\textwidth]{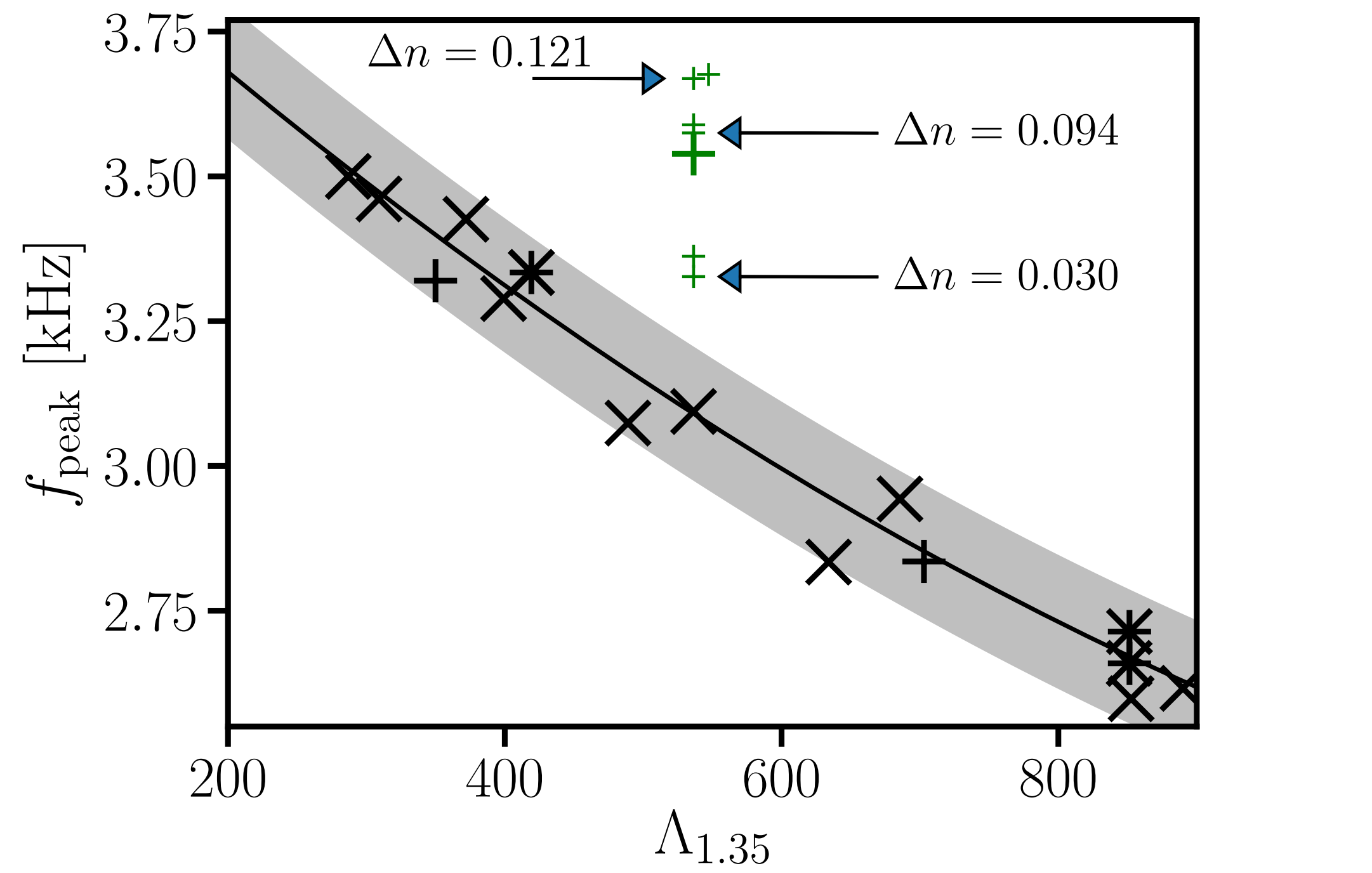}
	\caption{Dominant post-merger gravitational wave frequency $f_{\mathrm{peak}}$ as a function of tidal deformability $\Lambda_{1.35}$ for 1.35--1.35 $M_{\odot}$ binary neutron star mergers. 
		The DD2F-SF models with a first-order phase transition to deconfined quark matter (green symbols) appear as clear outliers (large symbol for DD2F-SF-1). 
		The solid black curve displays a least-squares fit for purely hadronic EOSs (including three models with hyperons marked by asterisks). 
		ALF2 and ALF4 EOSs are marked by black plus signs. 
		Arrows indicate DD2F-SF models 3, 6, and 7 with different density jumps $\Delta n$ (in fm$^{-3}$), but similar onset density and stiffness of quark matter. 
		The shaded gray band shows the $f_{\mathrm{peak}}$--$\Lambda$ relation uncertainty for hadronic EOSs.%
	}
	\label{fig4}
\end{figure}

While purely hadronic equations of state (EOSs) tend to follow a relatively tight correlation between these two quantities, the inclusion of a quark-hadron phase transition may lead to substantial deviations due to the softening of the EOS and the resulting changes in the remnant's compactness. This effect can serve as a strong observational signature of phase transitions in neutron star interiors. Figure~\ref{4} taken from \cite{ref1} illustrates this correlation and highlights how EOSs incorporating FOPTs deviate from the empirical trend established by hadronic models.

\section{Pulsar Timing Observations}
\label{4}
Pulsar timing provides a precise and independent method for constraining the neutron star equation of state (EOS), particularly through measurements of mass, spin, and long-term rotational evolution. Millisecond pulsars (MSPs), with their highly stable spin periods, allow for detailed tests of relativistic effects and interior composition. The presence of a first-order phase transition (FOPT) in neutron stars could introduce observable signatures in pulsar mass distributions, spin-down behavior, and glitch dynamics. Moreover, pulsar timing arrays (PTAs) offer an indirect method for probing phase transitions by detecting the stochastic gravitational wave background, which may contain imprints of early-universe phase transitions and binary neutron star mergers.

\subsection{Mass Measurements and Constraints on the Maximum Neutron Star Mass}

The discovery of neutron stars with masses exceeding \(2 M_{\odot}\) has imposed stringent constraints on EOS models. Traditional hadronic EOSs predict a maximum mass of around 2.1–2.3\(M_{\odot}\), but recent measurements suggest the existence of heavier neutron stars\cite{ref6}, possibly requiring the inclusion of an FOPT or exotic matter in their cores. The most massive confirmed neutron stars include PSR J0740+6620 (\(2.08 \pm 0.07 M_{\odot}\)) and PSR J0952-0607 (\(2.35 \pm 0.17 M_{\odot}\)), both of which challenge purely hadronic models. If a neutron star undergoes an FOPT, the transition to a quark core could lead to the twin-star phenomenon, where stars with nearly identical masses have significantly different radii.

The Tolman-Oppenheimer-Volkoff (TOV)\cite{ref13} equation governs the equilibrium structure of neutron stars:

\begin{equation}
\frac{dP}{dr} = - \frac{G}{c^2} \frac{(P + \epsilon)(M + 4\pi r^3 P/c^2)}{r^2 (1 - 2GM/rc^2)}
\end{equation}

where \(P\) is the pressure, \(\epsilon\) is the energy density, and \(M(r)\) is the enclosed mass. An FOPT would introduce a discontinuity in \(\epsilon\), leading to deviations from standard TOV predictions.

\begin{table}[h]
	\centering
	\caption{Maximum Neutron Star Mass Predictions for Different EOS Models}
	\begin{tabular}{lccc}
		\hline
		Pulsar Name & Measured Mass ($M_{\odot}$) & EOS Model & Max $M_{\text{NS}}$ ($M_{\odot}$) \\
		\hline
		PSR J0740+6620  & $2.08 \pm 0.07$  & APR4  & 2.2  \\
		PSR J0952-0607  & $2.35 \pm 0.17$  & SLy   & 2.1  \\
		PSR J0348+0432  & $2.01 \pm 0.04$  & DD2F-SF & 2.5  \\
		GW190814 Candidate & $2.59^{+0.08}_{-0.09}$  & MIT-Bag  & 2.6  \\
		\hline
	\end{tabular}
	\label{tab:max_mass}
\end{table}

Table~\ref{tab:max_mass} compares the maximum neutron star masses predicted by various equations of state (EOSs) with the measured masses of several observed compact objects\cite{ref16}. 
The first two pulsars, PSR~J0740$+$6620 and PSR~J0952$-$0607, exhibit high masses of $2.08 \pm 0.07\,M_\odot$ and $2.35 \pm 0.17\,M_\odot$, respectively, placing strong constraints on EOS stiffness. 
Notably, purely hadronic EOSs such as APR4 and SLy predict maximum masses of only $\sim$2.1--2.2\,$M_\odot$, which are barely sufficient—or even inadequate—to support the most massive observed pulsars. 
In contrast, EOS models that incorporate a first-order phase transition, such as DD2F-SF and MIT-Bag, can support significantly larger maximum masses ($2.5\,M_\odot$ and $2.6\,M_\odot$, respectively), making them better candidates for explaining both massive pulsars and the secondary object in GW190814. 
These results suggest that a phase transition may be necessary to reconcile theoretical EOS models with observations of ultra-massive neutron stars.

\subsection{Pulsar Spin Evolution and Evidence for Core Structure Changes}

The rotational evolution of pulsars is governed by magnetic dipole radiation and particle outflows, causing a gradual spin-down characterized by the braking index:

\begin{equation}
n = \frac{\nu \ddot{\nu}}{\dot{\nu}^2}
\end{equation}

where \(\nu\) is the spin frequency and \(\dot{\nu}\) is its time derivative.\cite{ref14} In the absence of an FOPT, neutron stars are expected to follow a smooth spin-down trajectory. However, if a phase transition occurs in the core, the moment of inertia \( I \) may change discontinuously, leading to observable anomalies in the pulsar’s spin-down rate.

Glitch events, where a pulsar suddenly increases in rotational frequency, may also provide evidence for phase transitions. In models where the neutron star core undergoes a transition to a quark phase, the redistribution of internal stresses could trigger starquakes or rapid reconfigurations of the superfluid component, resulting in measurable glitches. Observations of Vela pulsar glitches and their post-glitch recovery timescales suggest the involvement of complex core dynamics, potentially linked to a phase transition.

For the idealized magnetic dipole radiation mechanism, which represents the most commonly used standard spin-down model, the predicted braking index is $n = 3$.

\begin{table}[h]
	\centering
	\caption{Pulsar Braking Indices and EOS Predictions}
	\begin{tabular}{lccc}
		\hline
		Pulsar Name & Observed $ n $ & Standard EOS Prediction & EOS with FOPT? \\
		\hline
		PSR J0537-6910 & 1.6 & 3.0 (Dipole) & Yes \\
		PSR J0835-4510 (Vela) & 1.4 & 3.0 (Dipole) & Yes \\
		PSR B0540-69 & 2.1 & 3.0 (Dipole) & No \\
		\hline
	\end{tabular}
	\label{tab:braking_index}
\end{table}

As shown in Table~\ref{tab:braking_index}, the observed braking indices of several pulsars deviate significantly from the canonical value of 3, which is predicted by standard spin-down models assuming magnetic dipole radiation and purely hadronic equations of state (EOSs). Notably, PSR J0537--6910 and PSR J0835--4510 (Vela) exhibit braking indices well below 3, suggesting that additional physical processes may be at play. These deviations are consistent with scenarios involving a first-order phase transition (FOPT) in the neutron star core, which could lead to sudden changes in the moment of inertia. In contrast, PSR B0540--69 shows a braking index closer to the theoretical value, indicating no clear evidence for an FOPT in this case.

\subsection{Pulsar Timing Arrays and Gravitational Wave Background Constraints}

Pulsar timing arrays (PTAs) use highly stable millisecond pulsars to detect long-wavelength gravitational waves from astrophysical sources. These waves, originating from binary supermassive black holes and early-universe phase transitions, could also contain signatures of neutron star mergers where an FOPT occurs. The stochastic gravitational wave background measured by PTAs could reveal deviations from expected merger rates if a significant fraction of neutron stars collapse prematurely due to a phase transition.

The amplitude of the stochastic gravitational wave background (\(\Omega_{\text{GW}}\)) is given by:

\begin{equation}
\Omega_{\text{GW}}(f) = \frac{2\pi^2}{3H_0^2} f^3 S_h(f)
\end{equation}
where \( S_h(f) \) is the power spectral density of the strain and \( H_0 \) is the Hubble constant\cite{ref13}. If a first-order phase transition leads to an increased rate of prompt black hole formation in neutron star mergers, the detected background amplitude could differ from standard predictions.

Recent results from NANOGrav, EPTA, and SKA suggest possible deviations in the measured stochastic background, which may indicate unexpected merger dynamics or exotic EOS effects. If future PTA observations detect an enhanced GW background, it may suggest that an increased fraction of neutron star mergers undergo rapid collapse due to phase transitions, providing indirect evidence for deconfined quark matter in neutron stars.

\begin{table}[h]
	\centering
	\caption{PTA Constraints on the Gravitational Wave Background and EOS Implications}
	\begin{tabular}{lcc}
		\hline
		PTA Collaboration & GW Background Amplitude ($\Omega_{\text{GW}}$) & EOS Implication \\
		\hline
		NANOGrav 15-yr  & $10^{-8}$  & Marginally consistent with hadronic EOS \\
		EPTA            & $2 \times 10^{-8}$  & Possible FOPT effect \\
		SKA (Projected) & $10^{-9}$  & Strong constraints on EOS \\
		\hline
	\end{tabular}
	\label{tab:pta_constraints}
\end{table}
If future PTA observations detect an enhanced GW background, it may suggest that an increased fraction of neutron star mergers undergo rapid collapse due to phase transitions, providing indirect evidence for deconfined quark matter in neutron stars. Combining pulsar timing constraints with gravitational wave observations will help refine EOS models, offering a comprehensive approach to detecting first-order phase transitions in neutron stars.
\subsection{Comparison of Pulsar Timing Constraints and Gravitational Wave Observations}

While gravitational wave observations provide real-time insights into the structure of merging neutron stars, pulsar timing data offers long-term evolutionary constraints on EOS properties. A comparison of these two approaches is presented in Table \ref{tab:pta_gw_comparison}.

\begin{table}[h]
	\centering
	\caption{Comparison of Pulsar Timing and Gravitational Wave Constraints on the EOS}
	\begin{tabular}{lccc}
		\hline
		Method & Key Observable & Timescale & Sensitivity to FOPT \\
		\hline
		Pulsar Timing & Mass, Spin Evolution, Glitches & Years–Decades & Strong (Structural Changes) \\
		GW Inspiral & Tidal Deformability ($\Lambda$) & Seconds–Minutes & Moderate (NS Compactness) \\
		GW Post-Merger & $ f_{\text{peak}} $, Collapse Time & Milliseconds–Seconds & Strong (High-Density EOS) \\
		\hline
	\end{tabular}
	\label{tab:pta_gw_comparison}
\end{table}

The classification of sensitivity to a first-order phase transition (FOPT) is based on how directly each observational method probes the dense core conditions of neutron stars. Pulsar timing offers strong sensitivity due to its ability to detect long-term structural changes, such as variations in the moment of inertia and anomalies in spin evolution or glitch behavior, which may arise from a sudden phase transition in the stellar core. In contrast, gravitational wave signals from the inspiral phase provide only moderate sensitivity, as the tidal deformability parameter $\Lambda$ primarily reflects the neutron star's global compactness and may not significantly change unless the FOPT substantially softens the equation of state. Post-merger gravitational wave signals are strongly sensitive to FOPTs, as the remnant reaches extremely high densities where phase transitions are likely to occur. These transitions can lead to observable shifts in the dominant post-merger frequency ($f_{\mathrm{peak}}$) and rapid collapse timescales, offering direct insight into the high-density behavior of the equation of state.

The combination of these independent techniques provides a comprehensive strategy for detecting first-order phase transitions in neutron stars and constraining the nuclear EOS.

\section{Future Prospects}
\label{5}
The study of first-order phase transitions (FOPTs) in neutron stars is expected to advance significantly in the coming years due to improvements in observational capabilities, theoretical models, and computational simulations. While current observations provide strong constraints on the neutron star equation of state (EOS), the exact density at which deconfined quark matter may emerge, the nature of the phase transition, and its observational consequences are still under debate. Future advancements in observational techniques, theoretical modeling, and numerical simulations will be crucial in resolving these uncertainties.

Next-generation detectors such as the Einstein Telescope (ET) and Cosmic Explorer (CE) will significantly improve constraints on neutron star tidal deformability and post-merger oscillations\cite{ref16}, and the Laser Interferometer Space Antenna (LISA) is capable of detecting gravitational waves during the inspiral phase, helping distinguish between competing EoS models. In the future\cite{ref17}. The advancement of the TianQin project\cite{ref18} is expected to open a new window into low-frequency gravitational wave astronomy, enabling the detection of signals from neutron star binaries and phase transition events. Higher sensitivity in the kilohertz range will enable precise measurements of post-merger frequencies (\( f_{\text{peak}} \)), potentially revealing sudden EOS softening. A detectable suppression of post-merger signals due to quark core formation would provide strong evidence for a phase transition. Additionally, the stochastic gravitational wave background from neutron star mergers could show deviations if phase transitions significantly alter merger outcomes.

Future X-ray missions such as Athena and STROBE-X will refine neutron star radius measurements through high-precision spectroscopy and pulse profile modeling. Improved constraints on mass-radius relations will help distinguish between hadronic and hybrid EOSs. X-ray polarimetry from missions like eXTP may provide indirect evidence for phase transitions by revealing surface emission properties affected by a quark core. Long-term monitoring of QPOs and CRSFs could also identify shifts in neutron star compactness due to EOS modifications.

The Square Kilometer Array (SKA) will vastly improve neutron star mass measurements, particularly for ultra-massive objects exceeding \(2.5 M_{\odot}\), which would favor EOSs with phase transitions. High-precision pulsar timing will help detect glitch dynamics and moment of inertia changes, providing clues to core structure evolution. Pulsar timing arrays (PTAs) such as NANOGrav and EPTA will refine the detection of the stochastic gravitational wave background, which may contain signatures of neutron star mergers influenced by phase transitions.

Refining neutron star EOS models requires more sophisticated treatments of QCD at high densities. Functional renormalization group (FRG) methods and improved lattice QCD calculations will help constrain the transition density and latent heat of an FOPT. Bayesian inference combining gravitational wave and X-ray data will improve EOS constraints. General relativistic hydrodynamics simulations with advanced microphysics, including neutrino transport and magnetic field evolution, will offer better predictions for gravitational wave signatures of phase transitions.

A crucial test for phase transitions is the existence of twin-star configurations, where neutron stars of similar masses have distinct radii due to the presence of a quark core. Future mass-radius measurements from X-ray and gravitational wave observations could confirm this phenomenon. Additionally, optical and infrared follow-up observations of neutron star mergers may reveal kilonova light curve deviations, potentially indicating phase transition effects. The synergy between multi-messenger observations will be key to uncovering the nature of dense matter.

\section{Conclusion}
\label{6}
The study of first-order phase transitions (FOPTs) in neutron stars remains one of the most pressing challenges in modern astrophysics and nuclear physics. The presence of such a phase transition, leading to the emergence of a quark core or a mixed phase at high densities, has profound implications for the neutron star equation of state (EOS) and its observational signatures. This review has explored multiple approaches to identifying these transitions, including X-ray observations, gravitational wave detections, and pulsar timing measurements.

X-ray studies, particularly from NICER, XMM-Newton, and NuSTAR, provide precise constraints on neutron star mass-radius relations, thermal evolution, and accretion phenomena. Quasi-periodic oscillations (QPOs) and cyclotron resonance scattering features (CRSFs) offer indirect probes of neutron star compactness and interior composition. If a phase transition occurs, it could lead to measurable shifts in these observables, particularly in the form of sudden changes in QPO frequencies or variations in gravitational redshifts affecting CRSF energies.

Gravitational wave observations from LIGO, Virgo, and KAGRA have revolutionized the study of dense matter physics by providing direct constraints on neutron star deformability and post-merger dynamics. The tidal deformability parameter \(\Lambda\), extracted from the inspiral phase of events such as GW170817, strongly favors a relatively soft EOS, potentially hinting at a phase transition. Post-merger signals, though currently difficult to detect, could provide even stronger evidence by revealing sudden EOS softening due to quark core formation. The presence of ultra-massive compact objects, such as the 2.6\(M_{\odot}\) secondary in GW190814, suggests that an exotic EOS with a phase transition may be necessary to explain its existence if it is a neutron star rather than a black hole.

Pulsar timing offers another independent constraint on the EOS by measuring neutron star masses, spin evolution, and glitches. The discovery of neutron stars with masses exceeding \(2M_{\odot}\), such as PSR J0740+6620 and PSR J0952-0607, challenges purely hadronic EOSs and supports models incorporating an FOPT. Additionally, braking indices and glitch behavior in pulsars such as Vela suggest possible structural changes that could be linked to density-dependent phase transitions. Pulsar timing arrays (PTAs) provide further constraints by detecting the stochastic gravitational wave background, which may contain signatures of neutron star mergers influenced by phase transitions.

Theoretical models and numerical simulations have played a crucial role in predicting the effects of phase transitions in neutron stars. Hybrid EOSs incorporating a quark-hadron transition predict observable signatures in gravitational waves and mass-radius measurements, with some models suggesting the existence of twin-star configurations where stars of similar mass have distinctly different radii. Simulations of neutron star mergers incorporating phase transitions indicate that the presence of quark matter can significantly alter the post-merger gravitational wave spectrum, potentially reducing the dominant frequency \( f_{\text{peak}} \) and increasing the likelihood of rapid collapse.

Despite significant progress, no definitive observational proof of an FOPT in neutron stars has yet been obtained. However, current multi-messenger constraints place strong limits on the properties of such a transition, particularly its onset density and impact on neutron star stability. The combination of X-ray, gravitational wave, and pulsar timing observations will be essential in further refining these constraints, bringing us closer to confirming or ruling out the existence of deconfined quark matter in neutron stars.

\bibliographystyle{plain}

\end{document}